\documentclass[a4paper,11pt]{article}

\usepackage{latexsym}
\usepackage{epsfig}
\usepackage{color}
\usepackage{caption}
\def\R{ {\rm R \kern -.31cm I \kern .15cm}}
\def\C{ {\rm C \kern -.15cm \vrule width.5pt \kern .12cm}}
\def\Z{ {\rm Z \kern -.27cm \angle \kern .02cm}}
\def\N{ {\rm N \kern -.26cm \vrule width.4pt \kern .10cm}}
\def\1{{\rm 1\mskip-4.5mu l} }
\def\lsim{\raise0.3ex\hbox{$<$\kern-0.75em\raise-1.1ex\hbox{$\sim$}}}
\def\gsim{\raise0.3ex\hbox{$>$\kern-0.75em\raise-1.1ex\hbox{$\sim$}}}

\DeclareGraphicsExtensions{.eps}

\def\lsim{\raise0.3ex\hbox{$<$\kern-0.75em\raise-1.1ex\hbox{$\sim$}}}
\def\gsim{\raise0.3ex\hbox{$>$\kern-0.75em\raise-1.1ex\hbox{$\sim$}}}

\newcommand{\beq}{\begin{eqnarray}}
\newcommand{\eeq}{\end{eqnarray}}
\newcommand{\be}{\begin{eqnarray*}}
\newcommand{\ee}{\end{eqnarray*}}

\newcommand{\cc}{{c\bar{c}}}

\begin{document}

\title{{\bf Charmonium dissociation and recombination at LHC: Revisiting comovers}}

\vskip 8. truemm
\author{\bf 
E.G.~Ferreiro
}
\vskip 5. truemm

\date{}
\maketitle

\begin{center}
\small{
  Departamento de F{\'\i}sica de Part{\'\i}culas and IGFAE, Universidade de
  Santiago de Compostela, \\
  15782 Santiago de Compostela, Spain}
\end{center}
\vskip 5. truemm

\begin{abstract}
We present our results on 
charmonium production at the Large Hadron Collider energies within the
comover interaction model. The formalism 
includes both comover dissociation of $J/\psi$'s and 
possible secondary $J/\psi$ production through recombination. The
estimation of this effect is made without involving free parameters.
The comover interaction model also
incorporates an analytic treatment of initial-state nuclear shadowing.
With these
tools, the model successfully describes the centrality, transverse momentum and rapidity dependence of
the experimental data from PbPb collisions
at the LHC energy of $\sqrt{s} = 2.76$ TeV. We present predictions for PbPb collisions at $\sqrt{s} = 5.5$ TeV.
\end{abstract}
\vskip 3 truecm

\newpage

\section{Introduction}

Lattice QCD calculations predict that, at sufficiently large energy
densities, hadronic matter undergoes a phase transition to a plasma of
deconfined quarks and gluons (QGP).
Substantial
activity has been dedicated to
the research of high-energy heavy-ion collisions in order to reveal the
existence of this phase transition and to analyze the properties of strongly
interacting matter in the new phase.
The study of quarkonium production and
suppression is among the most interesting investigations in this field since,
in the
presence of a QGP, the charmonium yield would be further
suppressed due to color Debye screening \cite{Matsui86}.
Indeed, such an anomalous suppression
was first observed in PbPb collisions at top CERN SPS energy \cite{Alessandro05}.
Alternatively, the SPS experimental results could also
be described in terms of final state interactions of the $\cc$
pairs with the dense medium created in the collision, the so-called
comover interaction model (CIM) \cite{Capella97,Armesto98,Armesto99,Capella00}. This model does not assume thermal
equilibrium and, thus, does not use 
thermodynamical concepts.

The theoretical extrapolations to RHIC and LHC energies 
were led
mainly by two tendencies. On the one hand, the models that assume a deconfined phase
during the collision pointed out the growing importance of secondary
$J/\psi$ production due to regeneration of $\cc$ pairs in the plasma, the so-called {\it recombination}.
The total amount of $\cc$ pairs is created
in hard interactions during the early stages of the collision. Then,
either using kinetic theory and solving rate equations for the
subsequent dissociation and recombination of charmonium
\cite{Thews01,Grandchamp02,Yan06}, or assuming statistical coalescence at
freeze-out \cite{Braun00,Andronic03,Kostyuk03}, one obtains the final
$J/\psi$ yield. 
On the other
hand, the CIM with only
dissociation of $J/\psi$'s predicted \cite{Capella05} a stronger suppression at RHIC
than at SPS due to a larger density of produced soft particles in the
collision. It also predicted a stronger suppression at mid-rapidity --where the comover density is maximal-- than at forward rapidities.
Nevertheless, measurements of $J/\psi$ production in AuAu
collisions at $\sqrt{s} = 200$ GeV displayed
surprising results: the suppression at
mid-rapidity was on the same level as at SPS \cite{PHENIX07,Leitch07}.
Furthermore, the suppression at forward rapidity in AuAu collisions
was stronger than at mid-rapidity for the same collision energy. 

These facts demonstrated that in the CIM, which is based on the well-known gain and loss differential
equations in transport theory, the introduction of a recombination
term is actually required for a comprehensive adjustment \cite{Capella:2007jv}.
In the present work we use the updated version of the CIM \cite{Capella:2007jv}, that allows
recombination of $\cc$ pairs into secondary $J/\psi$'s. We will
estimate this effect by using the density of charm in proton-proton 
collisions at the 
same energy. 
Therefore, the model does not
involve any additional parameter. 
We will proceed as follows: 
In Section~\ref{sec:model} we remember the details of the model; 
the effects related to the initial-state shadowing and nuclear absorption, 
together with comover dissociation and regeneration are described.  
In Section~\ref{sec:lhc} we present our results for PbPb
collisions at the LHC energy of $\sqrt{s} = 2.76$ TeV and predictions at $\sqrt{s} = 5.5$ TeV.
Note that, at these energies, the CIM should not be considered to describe a final-state
interaction at the hadronic level. Indeed, at small values of the proper
time these comovers should be considered as a dense partonic medium.
Conclusions and final remarks
are given in Section~\ref{sec:conclusions}.

\section{Description of the model}\label{sec:model}
Let us briefly recall the main ingredients of the CIM.
The present version of the model contains an analytic treatment of initial-state nuclear
effects --the so-called {\it nuclear shadowing}--, together with the multiple scattering of the pre-resonant $c\bar{c}$ ̄
pair 
escaping the nuclear environment --the {\it nuclear absorption}--. The specific characteristics of the model are
the {\it interaction with the co-moving matter} and the {\it recombination} of $\cc$ into secondary
  $J/\psi$'s. 

The suppression of the $J/\psi$ is usually expressed through
the {\it nuclear modification factor}, $R^{J/\psi}_{AB}
(b)$, defined as the ratio of the $J/\psi$ yield in $AB$ and
{\it pp} scaled by the
number of binary nucleon-nucleon collisions, $n(b)$. We have then
\beq \label{eq:ratioJpsi}
R^{J/\psi}_{AB}(b) \;&=&\;
\frac{\mbox{d}N^{J/\psi}_{AB}/\mbox{d}y}{n(b)
  \,\mbox{d}N^{J/\psi}_{pp}/\mbox{d}y} \nonumber \\ 
\;&=&\; \frac{\int\mbox{d}^2s \, 
  \sigma_{AB}(b) \, n(b,s) \, S^{abs}(b,s) \, S_{J/\psi}^{sh}(b,s) \, S^{co}(b,s)
}{\int \mbox{d}^2 s \, \sigma_{AB} (b) \, n(b,s)} \;,
\eeq
where $\sigma_{AB}(b) = 1 - \exp [-\sigma_{pp}\, AB\, T_{AB}(b)]$, 
$T_{AB}(b) = \int\mbox{d}^2s T_A(s)T_B(b-s)$ is the nuclear 
overlap
function and $T_A(b)$ is 
the nuclear profile function,
obtained from Woods-Saxon nuclear densities \cite{Jager74}.

The number of
binary nucleon-nucleon collisions at impact parameter $b$, $n(b)$, is obtained upon integration over $\mbox{d}^2s$ of 
$n(b,s)$:
\beq
\label{eq:nbin}
n(b,s) \;=\; \sigma_{pp} AB \, T_A(s)\, T_B(b-s)/\sigma_{AB}(b)\;.
\eeq

The three additional factors in the numerator of
eq.~(\ref{eq:ratioJpsi}), $S^{abs}$, $S^{sh}$ and $S^{co}$, denote the
effects of nuclear absorption, shadowing and interaction with
the co-moving matter --both dissociation and recombination--, respectively. 
Any of these effects on $J/\psi$ production will lead to a deviation of $R^{J/\psi}_{AB}$ from unity.

It is commonly assumed that the nuclear absorption can be safely considered as negligible at the LHC \cite{Braun98,Capella:2006mb,Lourenco:2008sk} and thus
we will take $S^{abs}=1$ for the remainder of the discussion.

\subsection{Shadowing}

Coherence effects will lead to nuclear shadowing for both soft
and hard processes at high energy, and therefore also for the production of
heavy flavor. Shadowing can be calculated
within the Glauber-Gribov theory \cite{Gribov69} making use of
the generalized Schwimmer model of multiple scattering
\cite{Schwimmer75}. The second suppression factor in
eq.~(\ref{eq:ratioJpsi}) is then given by 
\beq
\label{eq:schwimmer}
S^{sh}(b,s,y) \;=\; \frac{1}{1 \,+\, A F(y_A) T_A(s)} \, \frac{1}{1
  \,+\, B F(y_B) T_B (b-s)} \;,
\label{eq:shad}
\eeq
where the function $F(y)$ encodes the dynamics of shadowing. 
Following the spirit
of the model presented in \cite{Capella99,Armesto03}, where
shadowing corrections are given without free parameters
in terms of the triple-Pomeron coupling determined from 
diffractive
data, one can write:
\beq
\label{eq22}
F(y)
=4\pi\int_{y_{min}}^{y_{max}}dy\ \frac{1}{\sigma_P(y)}\left.\frac{d\sigma^{PPP}}
{dydt}\right\vert_{t=0}
\;=\; G^{PPP} \left[\exp\left(\Delta \times y_{max} \right) \,-\, \exp
  \left(\Delta \times y_{min} \right) \right] / \Delta \;.
\eeq
It represents the coherence effects due to
the shadowing corrections expressed as the ratio of the triple-Pomeron cross section over the single-Pomeron exchange.
We take 
$y_{min} = \ln (R_A m_N/\sqrt{3})$
where 
$m_N$ is the nucleon mass and 
$R_A = (0.82~A^{1/3} + 0.58)$~fm
--the 
Gaussian nuclear radius.
The value of $y_{max}$ depends on the rapidity $y$
of the considered particle --the $J/\psi$ according to eq.~(\ref{eq:ratioJpsi})--,
the energy through 
$s_{NN}$
 and
the mass and the transverse momentum of the
produced particle through the transverse mass $m_T$, $y_{max} = \ln(s_{NN}/m_T^2)/2\pm y$ with the
$+(-)$ sign if 
the particle is produced in the hemisphere of nucleus $B(A)$. 
We have used $\Delta = 0.13$ and $G^{PPP} = 0.04$~fm$^2$ 
($G^{PPP}/\Delta = 0.31$ fm$^2$) corresponding to the Pomeron intercept $\alpha_P (0) = 1.13$.
The scale dependence of the shadowing appears in the expression of $y_{max}$, through
the transverse mass $m_T$. As a consequence, the shadowing corrections depend on the nature of the studied particle through its mass, and on its transverse momentum $p_T$.


Note that the above expression for the shadowing, eq.~(\ref{eq:shad}), can be applied to 
light and heavy particles, the difference between them coming from $y_{max}$ through the transverse mass $m_T$.

\begin{figure}[h!]
\vskip 0.3cm
\begin{minipage}[t]{.5\textwidth}
\begin{center}
\includegraphics[width=1.0\textwidth]{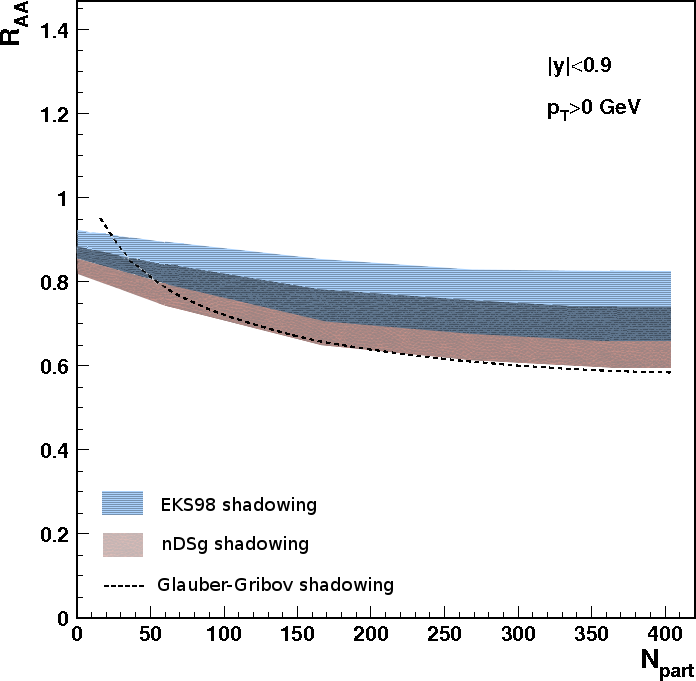}
\label{fig2a}
\end{center}
\end{minipage}
\begin{minipage}[t]{.5\textwidth}
\begin{center}
\includegraphics[width=1.0\textwidth]{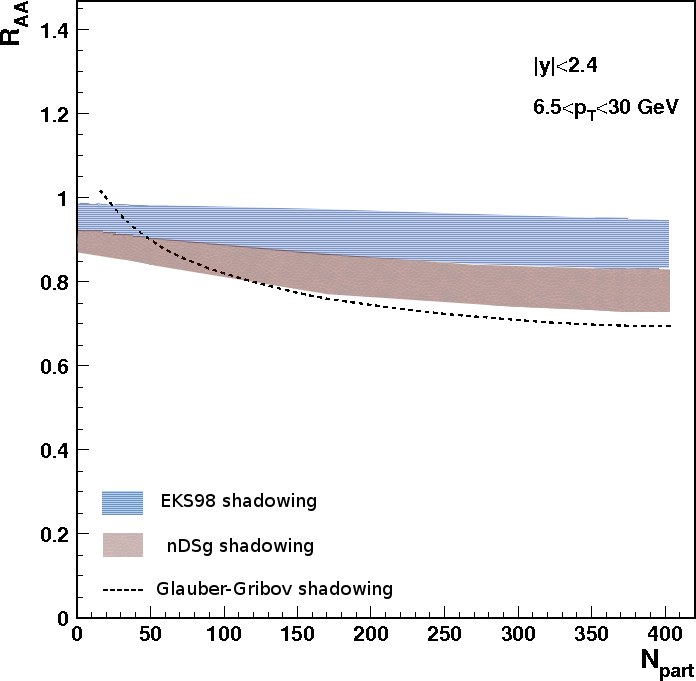}
\label{fig2b}
\end{center}
\end{minipage}
\vskip -0.4cm
\caption{(Color online) Centrality dependence of the Glauber-Gribov shadowing corrections for the $J/\psi$
compared to EKS98/nDSg calculations
performed according to \cite{Ferreiro:2008wc,Rakotozafindrabe:2011rw} in PbPb collisions at $\sqrt{s}=2.76~{\rm TeV}$.
The bands for the EKS98/nDSg models shown in the figure correspond to the uncertainty in the factorization scale.
This uncertainty has not been included in the Glauber-Gribov model, 
where the scale corresponds to the transverse mass.}
\label{fig2}
\hfill
\end{figure}

With the shadowing resulting from
the above equations a good description of the centrality
dependence of charged multiplicities is obtained both at RHIC \cite{Capella01} and LHC \cite{Capella:2011vi}
energies. 
Concerning heavy quarks, this shadowing 
roughly agrees with 
EKS98/nDSg predictions \cite{Ferreiro:2008wc,Rakotozafindrabe:2011rw,Ferreiro:2008td,Ferreiro:2009qr,Vogt:2010aa}.
In Fig.~\ref{fig2}
we show the comparison of the shadowing model applied here with some recent calculations developed within
the framework
of Refs. \cite{Ferreiro:2008wc,Rakotozafindrabe:2011rw}. One can see that the Glauber-Gribov inspired model agrees with the lowest band of nDSg according to \cite{Ferreiro:2008wc,Rakotozafindrabe:2011rw}.
In fact, our minimum bias shadowing is of the order of 0.66, to be compared to a minimal and maximal values of $0.65 \div 0.77$ for nDSg shadowing and $0.70 \div 0.84$ for EKS98.
Moreover, calculations of the shadowing developed at leading order 
within the framework of the Color Evaporation model \cite{Vogt:2010aa} lead in general to an slightly higher suppression than \cite{Ferreiro:2008wc}. Because of this, one can consider that the shadowing presented here agrees with the models mentioned above within uncertainties.

Note that while the particle production at SPS is dominated by low-energy effects,
i.e. nuclear absorption,
the RHIC domain already belongs to the high-energy
regime, where nuclear shadowing becomes relevant, and the
combined effect of shadowing and energy-momentum conservation
should be accounted for at forward
rapidities. At LHC, shadowing is expected to be strong
while nuclear absorption is a small effect that can be
neglected. We will now proceed with the
discussion of the specific comover-interaction effects.

\subsection{Dissociation by comover interaction and recombination}
The CIM was originally developed in the nineties in order to explain both the
suppression of charmonium yields
\cite{Capella97,Armesto98,Armesto99,Capella00,Capella05,Brodsky88,Koch90}
and the strangeness enhancement \cite{Capella95,Capella96} in
nucleus-nucleus collisions at the SPS. At those energies, where the
recombination effects are negligible, the rate equation governing the
density of charmonium in the final state, $N_{J/\psi}$, can be written
in a simple form assuming a pure longitudinal expansion of the system and boost invariance.
The
density of $J/\psi$ at a given 
transverse coordinate $s$, impact parameter $b$ and rapidity $y$ is then given by
\beq
\label{eq:comovrateeq}
\tau \frac{\mbox{d} N_{J/\psi}}{\mbox{d} \tau} \, \left( b,s,y \right)
\;=\; -\sigma_{co} N^{co}(b,s,y) N_{J/\psi}(b,s,y) \;,
\eeq
where $\sigma_{co}$ is the cross section of charmonium dissociation
due to interactions with the co-moving medium, with density $N^{co}$. It 
was fixed 
from fits to low-energy experimental data to be $\sigma_{co} = 0.65$~mb~\cite{Armesto99}.

In order to
incorporate the effects of recombination, one has to include an
additional gain term proportional to the squared density of open charm
produced in the collision. Eq.~(\ref{eq:comovrateeq}) is then
generalized to
\beq
\label{eq:recorateeq}
\tau \frac{\mbox{d} N_{J/\psi}}{\mbox{d} \tau} \, \left( b,s,y \right)
\;=\; -\sigma_{co} \left[ N^{co}(b,s,y) N_{J/\psi}(b,s,y) \,-\,
  N_c(b,s,y) N_{\bar{c}} (b,s,y) \right] \;,
\eeq
where we have assumed that the effective recombination cross section
is equal to the dissociation cross section. Note that these two
  cross sections have to be similar but not necessarily equal. We have
  taken the simplest possibility. Therefore, the extension of the model conducing to 
include recombination 
does not involve additional parameters\footnote{Strictly speaking,
the equivalence between breakup and recombination cross sections 
only holds if one consider the direct $J/\psi$ production.
Considering the feed down in a detailed way can induce differences between both interaction cross sections.}.
All the densities involved in eq.~(\ref{eq:recorateeq}) are assumed to decrease as $1/\tau$.
The approximate solution of eq.~(\ref{eq:recorateeq}) is given by
\beq
\label{eq:fullsupp}
S^{co}(b,s,y) \;=\; \exp \left\{-\sigma_{co}
  \,\left[N^{co}(b,s,y)\,-\, \frac{N_c(b,s,y)
  N_{\bar{c}} (b,s,y)}{N_{J/\psi}(b,s,y)} \right] \, \ln
\left[\frac{N^{co}(b,s,y)}{N_{pp} (0)}\right] \right\} \;,
\eeq
where the first term in the exponent corresponds 
to the exact solution of eq.~(\ref{eq:comovrateeq}),
i.e. the
survival probability of a $J/\psi$ interacting with comovers
\cite{Capella05}.
The breakup and recombination in the above equation do not need to occur on the same time.
The density of comovers is calculated following the same lines as in 
\cite{Capella:2011vi} together with the shadowing correction:
\beq
\label{eq:DensComov}
N^{co} (b,s,y) \;=\; N^{co}_{NS} (b,s,y) \, S^{sh}_{ch} (b,s,y) \;,
\eeq
where $S_{ch}^{sh}$ denotes the shadowing for light particles,
calculated according to eq.~(\ref{eq:shad}). 
The non-shadowed (NS) multiplicity of comovers is taken as proportional to the number of 
nucleon-nucleon collision according to eq.~(\ref{eq:nbin}) 
\beq
N^{co}_{NS} (b,s,y) \;=\; N^{pp}(b,s,y) \, n(b,s) \; ,
\eeq
where $N^{pp}$ represents the comover density in $pp$ collisions, essentially
$N^{pp} (y) = \frac{3}{2} (\mbox{d} N^{ch}/\mbox{d} y  )^{pp}$.
In fact, when using at mid-rapidity the value 
$(\mbox{d}
N^{ch}/\mbox{d} \eta  )^{pp}_{y=0}=3.8$, i.e. the
inelastic value of $d N^{pp}/d\eta$ at $\sqrt{s}=2.76$ TeV, 
a good agreement 
with experimental data on charged particle multiplicities is obtained 
\cite{Capella:2011vi}.

The density of open and hidden 
charm in $AA$ collisions, $N_c,N_{\bar{c}}$ and $N_{J/\psi}$,
respectively, can be computed from their densities in {\it
  pp} collisions as $N_c^{AA} (b,s) = n(b,s)
S_{HQ}^{sh}(b,s)N_c^{pp}$, with similar expression for
$N_{\overline{c}}^{AA}$ and $N_{J/\psi}^{AA}$. Here $n(b,s)$ is 
given by eq.~(\ref{eq:nbin}) and $S_{HQ}^{sh}$ is the shadowing factor
for heavy quark production, given by eq.~(\ref{eq:shad}).
Eq.~(\ref{eq:fullsupp}) becomes
\beq
S^{co}(b,s,y) \;=\; \exp \left\{-\sigma_{co} \,\left[N^{co}(b,s,y)- C(y)
    n(b,s)S_{HQ}^{sh}(b,s) \right] \, \ln
  \left[\frac{N^{co}(b,s,y)}{N_{pp} (0)}\right] \right\}
\eeq
where
\beq
\label{eq:Cratio}
C (y) 
\;=\; \frac{\left(\mbox{d}N^{c}_{pp}/\mbox{d}y
  \right)\left(\mbox{d}N^{\bar{c}}_{pp}/\mbox{d}y
  \right)}{\mbox{d}
  N^{J/\psi}_{pp}/\mbox{d} y} 
\;=\; \frac{\left(\mbox{d}N^{\cc}_{pp}/\mbox{d}y
  \right)^2}{\mbox{d} 
  N^{J/\psi}_{pp}/\mbox{d} y} \;=\; \frac{\left(\mbox{d}
    \sigma^{\cc}_{pp}/\mbox{d}y 
  \right)^2}{\sigma_{pp} \,
  \mbox{d}\sigma^{J/\psi}_{pp}/\mbox{d} y} \;.
\eeq
The quantities in the rightmost term in eq.~(\ref{eq:Cratio}) are all
related to {\it pp} collisions at the corresponding energy. The value for 
$\mbox{d}\sigma^{J/\psi}_{pp}/\mbox{d} y$
can be 
taken from experimental data \cite{Abelev:2012kr}
or from a model for extrapolation of the
experimental results \cite{Bossu:2011qe}.
The $\cc$ pairs are mostly in charmed mesons,
such as $D$ and $D^*$ and the corresponding values could be extracted from the 
experiment \cite{:2012sx}, leading to an estimation of $\mbox{d}\sigma^{\cc}_{pp}/\mbox{d}y$,
as we will discuss below.
For $\sigma_{pp}$ 
we use the 
non-diffractive value
$\sigma_{pp}$=54 mb \cite{Capella:2011vi} 
at $\sqrt{s_{NN}}= 2.76$~TeV.

With $\sigma_{co}$ fixed from experiments at low energy, where
recombination effects are negligible, the model, formulated above,
should be self-consistent at high energies. Note, however,
  that $\sigma_{co}$ could change when the energy increases. We do not
  expect this effect to be important and, since we are unable to
  evaluate the magnitude of this eventual change, we have used the same
  value $\sigma_{co} = 0.65$~mb at all energies. 

\section{Results}\label{sec:lhc}

\begin{figure}[h!]
\vskip 0.3cm
\begin{minipage}[t]{.5\textwidth}
\begin{center}
\includegraphics[width=1.0\textwidth]{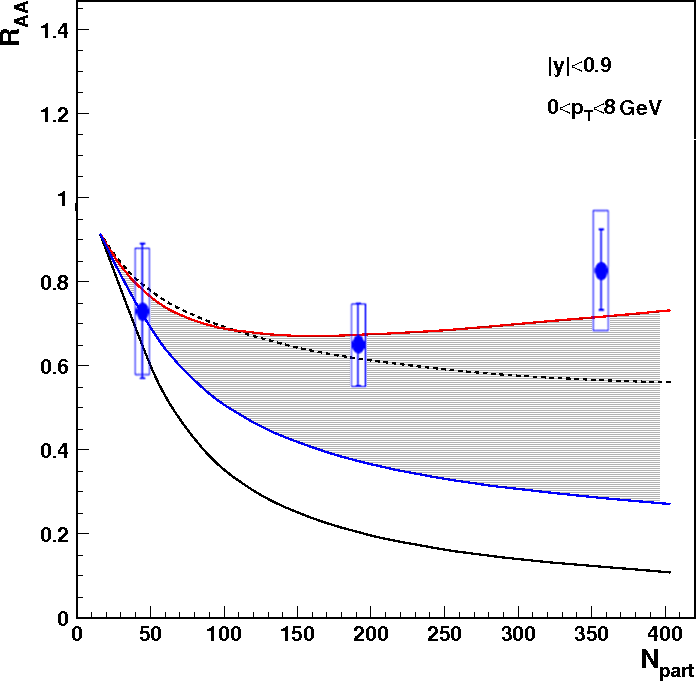}
\label{fig1a}
\end{center}
\end{minipage}
\begin{minipage}[t]{.5\textwidth}
\begin{center}
\includegraphics[width=1.0\textwidth]{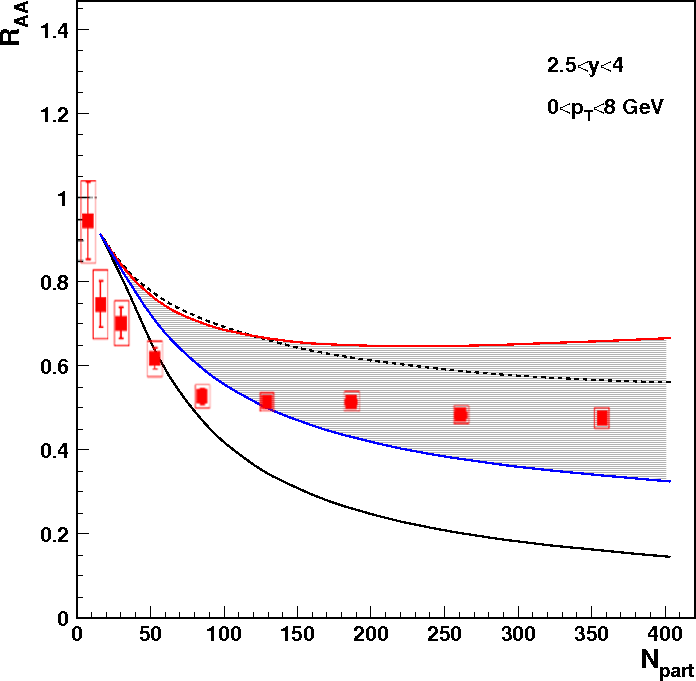}
\label{fig1b}
\end{center}
\end{minipage}
\vskip -0.4cm
\caption{(Color online) Results on the centrality dependence of
 the $J/\psi$ nuclear modification factor in PbPb collisions at 2.76 TeV at mid (left) and forward (right)
rapidity
compared to ALICE data \cite{Abelev:2012rv,Wiechula:2012mh,ScomparinQM2012,ArnaldiQM2012}. 
The dashed line corresponds to the {\it shadowing} effect on the $J/\psi$.
The lowest continuous line (black) corresponds to the combined effect of the {\it shadowing} and the {\it comover dissociation}.
The shadowed area corresponds to our result when the {\it shadowing}, the {\it comover dissociation} and the {\it recombination} are taken into account.
The uncertainty takes into account the variation between the minimum (blue line) and maximal (red line) values of $C(y)$.}
\label{fig1}
\hfill
\end{figure}

Our results for the centrality dependence of the $J/\psi$ nuclear modification factor in PbPb collisions at 
2.76 TeV are presented in Fig.~\ref{fig1} 
compared to ALICE experimental data \cite{Abelev:2012rv,Wiechula:2012mh,ScomparinQM2012,ArnaldiQM2012} at mid and forward rapidities.
The different
contributions to $J/\psi$ suppression are shown. 
Note that 
the initial-state effect is just the
shadowing, which can induce a suppression of $R_{AA}=0.6$ for the more central collisions \cite{Rakotozafindrabe:2011rw}.
The combined effect of shadowing and comover dissociation
 gives a too strong suppression
compared to experimental data. 
We therefore proceed
to estimate the effect of recombination.

For the charmonium cross section $pp$ measurements around mid-rapidity are available
from ALICE \cite{Abelev:2012kr} at 2.76 TeV which corresponds to $\frac{d\sigma_{pp}^{J/\psi}}{dy}= 3.73$ $\mu$b.
We consider that realistic values of $C(y)$ at mid-rapidity at 2.76 TeV are in the range of a
minimum value of 2
up to a maximal value of 3
which corresponds to a cross section $\frac{d\sigma_{pp}^{c\bar{c}}}{dy} \approx 0.6 \div 0.8$ mb.
This agrees with the estimated values in \cite{Zhao:2011cv},
and corresponds to a $\sigma^{tot}_{c\bar{c}}$ around 5 mb, which agrees well with experimental data \cite{:2012sx}.
These values are higher than the ones reported in \cite{Andronic:2011yq}, where data is also reproduced.
Note nevertheless that there is no contradiction, since in \cite{Andronic:2011yq} the initial-state
shadowing is not considered. 
This shadowing, that affects also heavy flavors, would imply an extra suppression leading naturally to the choice of higher input charm cross sections 
$\mbox{d}\sigma^{\cc}_{pp}/\mbox{d}y$.
In other words, one can consider that in the approach developed in \cite{Andronic:2011yq} the choice
of smaller $\frac{d\sigma_{pp}^{c\bar{c}}}{dy}$, $\frac{d\sigma_{pp}^{c\bar{c}}}{dy} \approx 0.3 \div 0.4$ mb, takes into account an effect of shadowing that reduces the
input charm
cross section up to $1/2$.

We expect the effect of recombination to be stronger at
mid than at forward rapidities. At $y \neq 0$ the recombination term is smaller
since the rapidity distribution of $D$, $D^*$ is
narrower than the one of comovers, which induces a decrease of the $C(y)$-value.
This will produce a decrease of $R_{AA}^{J/\psi}$ with increasing $y$. Note 
nevertheless that this effect may be 
compensated by the increase of $R_{AA}^{J/\psi}$ due to a smaller density of
comovers at $y \neq 0$, which induces less dissociation.
We have chosen a smooth behavior of $C(y)$ with rapidity, 
which in the rapidity range $2.5<y<4$ corresponds to a mean reduction in the $C(y)$-value of the order of 20\%. 
Taking $\frac{d\sigma_{pp}^{J/\psi}}{dy}$ in this forward rapidity range, $\frac{d\sigma_{pp}^{J/\psi}}{dy} = 2.23$ $\mu$b \cite{Abelev:2012kr},
these $C(y)$-values correspond 
to an input charm cross section $\frac{d\sigma_{pp}^{c\bar{c}}}{dy} \approx 0.4 \div 0.6$ mb in the forward rapidity region.
Our procedure gives a reasonable description of data when recombination is taken into account.

\noindent Some considerations 
apply here:

First, the behavior of the nuclear modification factor $R_{AA}^{J/\psi}$ for different rapidity ranges
changes depending of the amount of recombination considered. 
If the recombination $C(y)$-value is considered to be small --lower limit--, 
the total suppression will be controlled essentially by the comover dissociation,
and the nuclear modification factor will follow the same behavior as the usual comover
suppression, but with a higher absolute value. 
On the other hand, when the $C(y)$-value is taken as its upper limit,
the recombination term controls the total suppression,
inducing a decrease of the 
nuclear modification factor when going to forward rapidities.

Second, the shadowing corrections for heavy quarks we use here have been calculated in a very simple way, 
according to eq.~(\ref{eq:shad}) within the Glauber-Gribov theory. The advantage is that all calculations can be 
easily done analytically. 
This shadowing 
roughly agrees with EKS98/nDSg predictions, as shown in Fig.~\ref{fig2}, in particular in the mid-rapidity region.
Nevertheless, the Glauber-Gribov shadowing is almost constant with rapidity.
Even if the $y$-dependence of the different shadowing models in $AA$ collisions and in the rapidity ranges here considered
is quite flat, as it shown in \cite{Abelev:2012rv} and Refs. \cite{Rakotozafindrabe:2011rw,Vogt:2010aa} therein,
we are aware that our approach can induce an small overestimation of the shadowing suppression in the forward rapidity region.
In fact we contemplate the upgrading of the CIM by the introduction of EKS98/nDSg/EPS09 according to \cite{Rakotozafindrabe:2011rw,Ferreiro:2008wc,Ferreiro:2013pua}.\\

We proceed now to study the behavior of the nuclear modification factor when different transverse momentum cuts are introduced. 
In order to avoid an unnecessary complication of the notation, the explicit $p_T$ dependence is not shown in our equations. The $p_T$ comover distribution follows essentially the lines of reference \cite{Capella:2006fw}.
Moreover, the $p_T$ dependence also enter through
the shadowing --both on the $J/\psi$ and the comovers--.
The main effect would be smaller shadowing suppression for the comovers a high $p_T$,
which leads to stronger $J/\psi$ suppression at high $p_T$ due to comover interaction.
This effect is not compensate by the smaller shadowing corrections that affect the $J/\psi$,
since the higher mass of the $c\bar{c}$ pair makes the shadowing difference between the low and high $p_T$ regions less important that for the case of the comovers.
A detail work will be developed in \cite{CapFerr2014}.

\begin{figure}[h!]
\vskip 0.75cm
\begin{minipage}[t]{.5\textwidth}
\begin{center}
\includegraphics[width=1.0\textwidth]{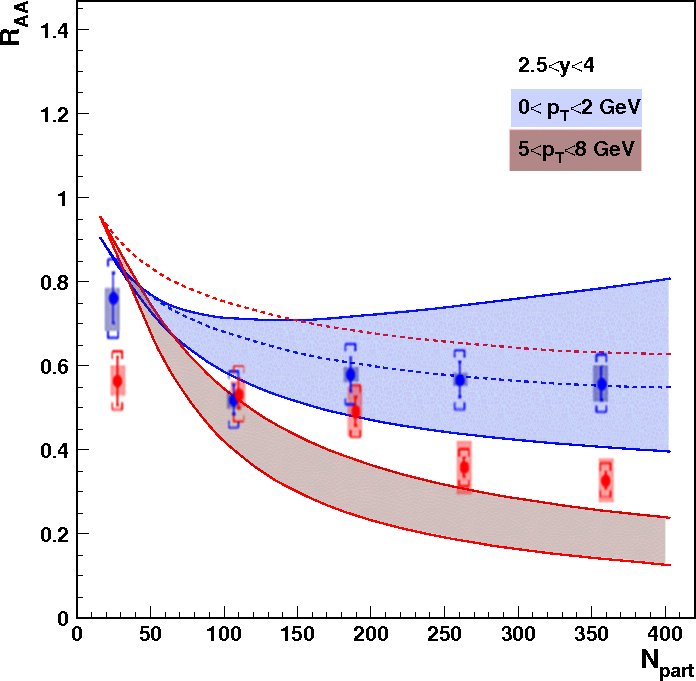}
\label{fig3a}
\end{center}
\end{minipage}
\begin{minipage}[t]{.5\textwidth}
\begin{center}
\includegraphics[width=1.0\textwidth]{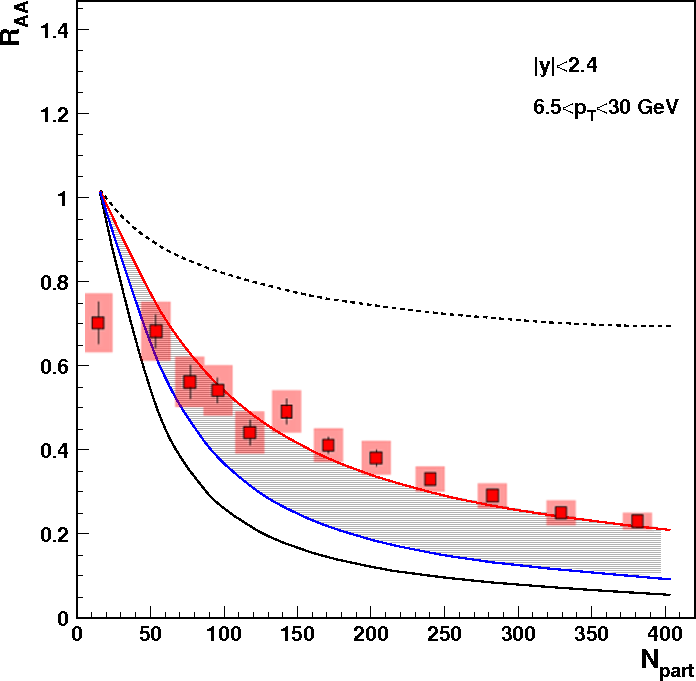}
\label{fig3b}
\end{center}
\end{minipage}
\vskip -0.3cm
\caption{(Color online) Results on the centrality dependence of
the $J/\psi$ nuclear modification factor in PbPb collisions at 2.76 TeV
compared to ALICE forward rapidity data \cite{ArnaldiQM2012} in the $p_T$ ranges $0<p_T<2$ GeV and $5<p_T<8$ GeV (left)
and compared to CMS mid-rapidity data \cite{Chatrchyan:2012np,MironovQM2012,MoonQM2012} in the $p_T$ range $6.5<p_T<30$ GeV (right).
The dashed line corresponds to the {\it shadowing} effect on the $J/\psi$.
The lowest continuous line (black) in the right figure
corresponds to the combined effect of the {\it shadowing} and the {\it comover dissociation}.
The shadowed areas correspond to our result when the {\it shadowing}, the {\it comover dissociation} and the {\it recombination} are taken into account.
The uncertainty takes into account the variation between the minimum and maximal values of $C(y)$.}
\label{fig3}
\hfill
\end{figure}

In  Fig.~\ref{fig3} (left)
our results for 2 different $p_T$ ranges, $0<p_T<2$ GeV and $5<p_T<8$ GeV, are compared to ALICE forward rapidity data \cite{ArnaldiQM2012}.
Clearly, the amount of recombination is more important in the low $p_T$ region. 
The comparison to CMS data at mid-rapidity and higher $p_T$, $6.5<p_T<30$ GeV \cite{Chatrchyan:2012np,MironovQM2012,MoonQM2012}, 
in Fig.~\ref{fig3} (right) emphasizes this finding: 
the amount of recombination is much smaller than in the mid-rapidity range at low $p_T$, as shown in Fig.~\ref{fig1} (left). It is important 
to point out 
that this experimental fact --stronger suppression at higher $p_T$-- cannot be due to initial-state shadowing effects on the $J/\psi$: 
the presence of shadowing corrections, more relevant at lower $p_T$, 
acts
in the opposite direction. 

\begin{figure}[h!]
\vskip 0.5cm
\begin{minipage}[t]{.5\textwidth} 
\begin{center}
\includegraphics[width=1.0\textwidth]{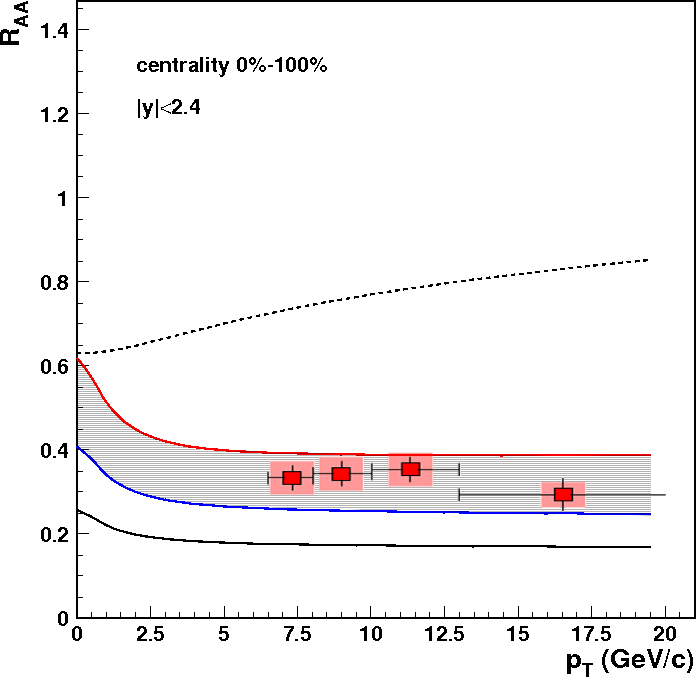}
\label{fig4a} 
\end{center} 
\end{minipage}
\begin{minipage}[t]{.5\textwidth}
\begin{center}
\includegraphics[width=1.0\textwidth]{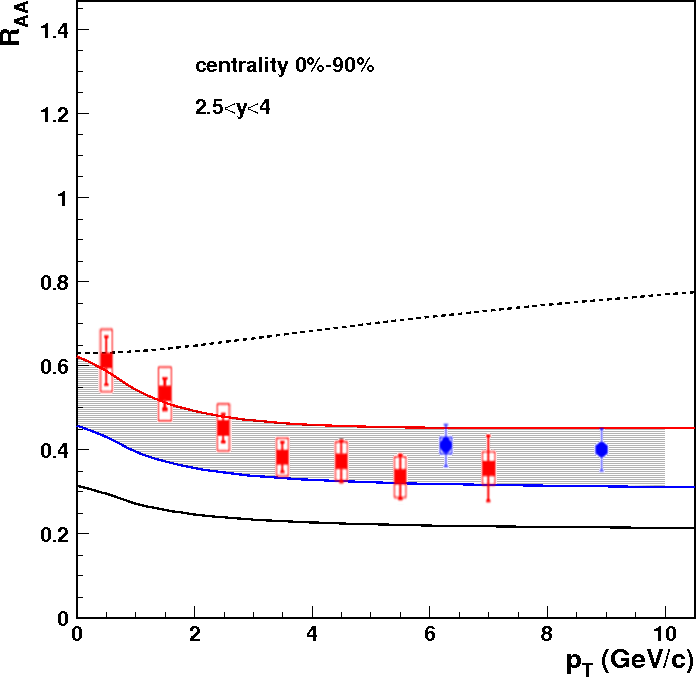}
\label{fig4b}
\end{center}
\end{minipage}
\vskip -0.3cm
\caption{(Color online) Results on the transverse dependence of
the $J/\psi$ nuclear modification factor in PbPb collisions at 2.76 TeV
compared to CMS data \cite{Chatrchyan:2012np,MironovQM2012} at mid-rapidity (left) and to ALICE \cite{ArnaldiQM2012} and CMS data \cite{Chatrchyan:2012np} at forward rapidity (right). The CMS data on the left part of the figure corresponds to the rapidity range $1.6 < |y| < 2.4$, 
while the ALICE data lie in the range $2.5 < y < 4$.
The dashed line corresponds to the {\it shadowing} effect on the $J/\psi$.
The lowest continuous line (black) corresponds to the combined effect of the {\it shadowing} and the {\it comover dissociation}.
The shadowed area corresponds to our result when the {\it shadowing}, the {\it comover dissociation} and the {\it recombination} are taken into account.
The uncertainty takes into account the variation between the minimum (blue line) and maximal (red line) values of $C(y)$.}
\label{fig4}
\hfill 
\end{figure}

Moreover, our results on the dependence of the $J/\psi$ nuclear modification factor on the transverse momentum are shown
in Fig.~\ref{fig4}
compared to CMS \cite{Chatrchyan:2012np,MironovQM2012} and
ALICE \cite{ScomparinQM2012} data.
An overall agreement is obtained.

\begin{figure}[h!]
\vskip 0.5cm
\begin{minipage}[t]{.5\textwidth}
\begin{center}
\includegraphics[width=1.0\textwidth]{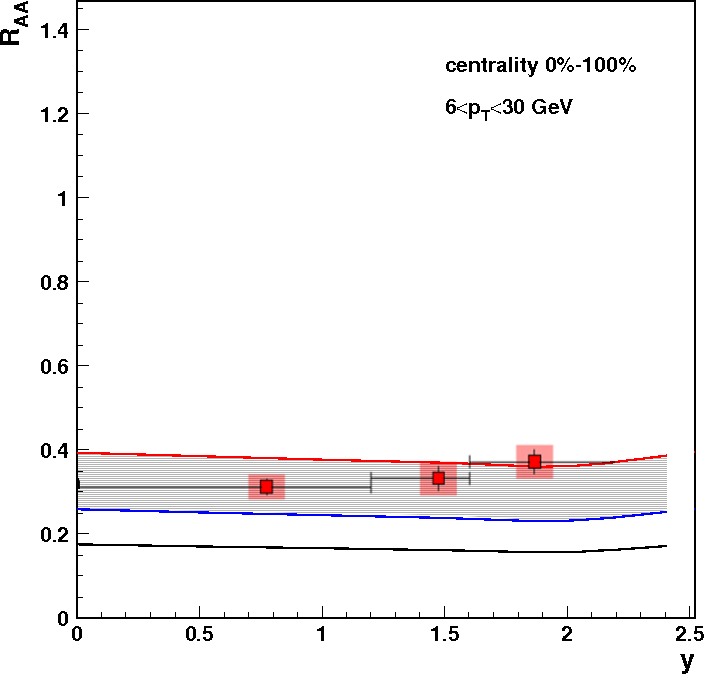}
\label{fig5a}
\end{center}
\end{minipage}
\begin{minipage}[t]{.5\textwidth}
\begin{center}
\includegraphics[width=1.0\textwidth]{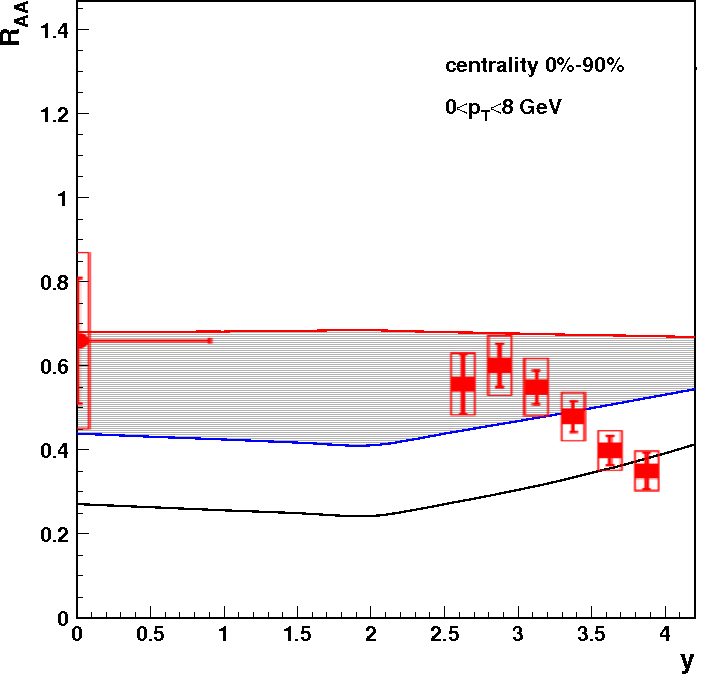}
\label{fig5b}
\end{center}
\end{minipage}
\vskip -0.3cm
\caption{(Color online) Results on the rapidity dependence of
the $J/\psi$ nuclear modification factor in PbPb collisions at 2.76 TeV
compared to CMS data \cite{Chatrchyan:2012np,MironovQM2012} at mid-rapidity (left) and to ALICE \cite{ScomparinQM2012} 
at mid and forward rapidity (right). 
The dashed line corresponds to the {\it shadowing} effect on the $J/\psi$.
The lowest continuous line (black) corresponds to the combined effect of the {\it shadowing} and the {\it comover dissociation}.
The shadowed area corresponds to our result when the {\it shadowing}, the {\it comover dissociation} and the {\it recombination} are taken into account.
The uncertainty takes into account the variation between the minimum (blue line) and maximal (red line) values of $C(y)$.}
\label{fig5}
\hfill
\end{figure}

We continue by showing our results versus rapidity. 
In Fig.~\ref{fig5}
(left) we compare our results with CMS data \cite{Chatrchyan:2012np,MironovQM2012} in the mid-rapidity region. An overall agreement is obtained.
The fact that we do not reproduce the detailed behavior of ALICE data \cite{ScomparinQM2012}, as can be seen in Fig.~\ref{fig5} (right), 
is extremely instructive. 
As we have mentioned above, 
we have chosen a very conservative behavior of $C(y)$ with rapidity, which leads to 
a mean reduction in the $C(y)$-value of the order of 20\% in the rapidity range $2.5<y<4$. 
The data in this region agrees with this choice
in the interval $2.5<y<3.5$, while for the most forward points an scenario with smaller amount of recombination or only dissociation without recombination is clearly favored.
Note that the lower curve in the right-hand side of Fig.~\ref{fig5}
corresponds to the behavior of the comover dissociation,
whose amount decreases rapidly when going from 2 to 4 in the rapidity range. More realistic $y$-dependent value of $C(y)$ could modify these curves.

We finish by presenting our predictions for the LHC energy of 5.5 TeV. 
Here we have let vary our $C(y)$-value between 2 and 4.
Taking $B R_{ll} \times \frac{d\sigma_{pp}^{J/\psi}}{dy} |_{y=0} \approx 350$ nb \cite{Bossu:2011qe}
and $\sigma_{pp} = 60$ mb,
these $C(y)$-values correspond
to an input charm cross section $\frac{d\sigma_{pp}^{c\bar{c}}}{dy} \approx 0.8 \div 1.15$ mb in the mid-rapidity region.
In the rapidity range $2.5<y<4$ we take a mean reduction in the $C(y)$-value of the order of 20\%, 
which corresponds to an input charm cross section $\frac{d\sigma_{pp}^{c\bar{c}}}{dy} \approx 0.6 \div 0.9$ mb when 
$B R_{ll} \times \frac{d\sigma_{pp}^{J/\psi}}{dy} \approx 250$ nb \cite{Bossu:2011qe}. 
\begin{figure}[h!]
\vskip 0.5cm
\begin{minipage}[t]{.5\textwidth}
\begin{center}
\includegraphics[width=1.0\textwidth]{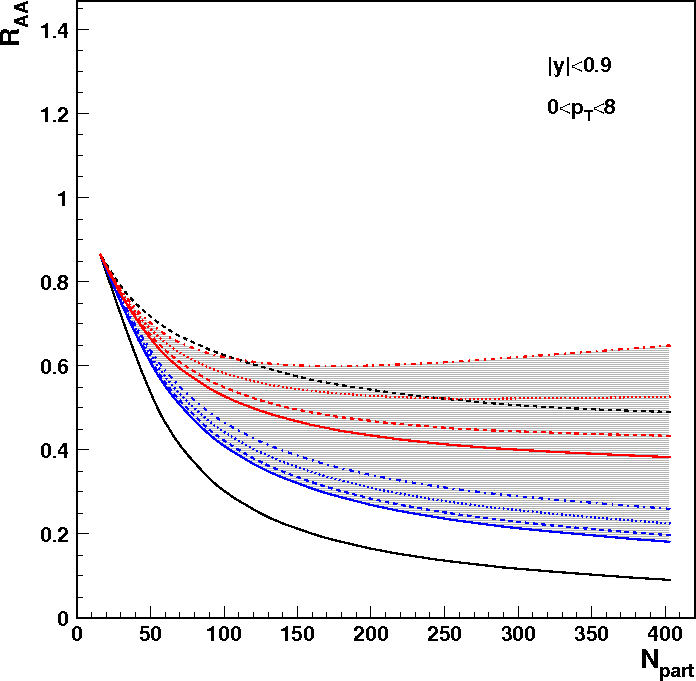}
\label{fig6a}
\end{center}
\end{minipage}
\begin{minipage}[t]{.5\textwidth}
\begin{center}
\includegraphics[width=1.0\textwidth]{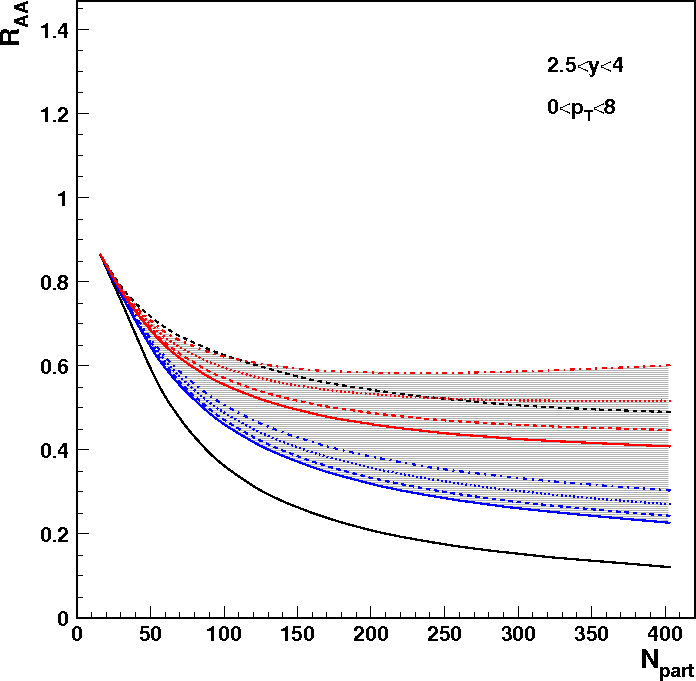}
\label{fig6b}
\end{center}
\end{minipage}
\vskip -0.3cm
\caption{(Color online) Results on the centrality dependence of
 the $J/\psi$ nuclear modification factor in PbPb collisions at 5.5 TeV at mid (left) and forward (right)
rapidity.
The dashed black line corresponds to the {\it shadowing} effect on the $J/\psi$.
The lowest continuous line (black) corresponds to the combined effect of the {\it shadowing} and the {\it comover dissociation}.
The shadowed area corresponds to our result when the {\it shadowing}, the {\it comover dissociation} and the {\it recombination} are taken into account.
The uncertainty takes into account the variation between the minimum (blue line) and maximal (red line) values of $C(y)$.
For the mid-rapidity region, we have taken, from down to up, $\frac{d\sigma_{pp}^{c\bar{c}}}{dy}=0.8, 0.85, 0.9, 0.95, 1.0, 1.1, 1.15$ mb
and for the forward rapidity region $\frac{d\sigma_{pp}^{c\bar{c}}}{dy}=0.65, 0.67, 0.71, 0.74, 0.82, 0.84, 0.87, 0.89$ mb.}
\label{fig6}
\hfill
\end{figure}
Our results in the mid and forward rapidity ranges are plotted in Fig.~\ref{fig6}

Note that in the previous work \cite{Capella:2007jv} 
stronger shadowing corrections --of around 20\% larger than the present ones-- for the heavy quark production were considered, 
which implied both less recombination effects and
stronger total suppression. 
This demonstrates that the implementation of the initial-state effects 
is more relevant than what is currently assumed in the recombination approaches, since it affects 
the probability of regeneration. 

\section{Conclusion}\label{sec:conclusions}

In this work we have studied the combined effect of $J/\psi$ dissociation and
recombination of $\cc$ pairs into $J/\psi$ in the comover interaction
model. 
This model does not assume thermal equilibrium of the matter
produced in the collision and
includes a comprehensive treatment of initial-state effects, such as
shadowing.
We estimate the magnitude of the recombination term from
$J/\psi$ and open charm yields in  {\it
  pp} collisions at LHC.
Without any adjustable parameters, the centrality, transverse momentum and
rapidity dependence of experimental data is reproduced.

In our approach, the magnitude of the
recombination effect is 
controlled by the total charm cross section in ${\it pp}$ collisions.
Note that, contrary to the results in \cite{Capella:2007jv}, 
where the combined effect of initial-state shadowing and comover dissociation appeared to overcome the effect
of parton recombination at 5.5 TeV, we find here that 
the recombination 
effects are of crucial importance in PbPb collisions at LHC energies and 
 can dominate over the suppression, 
in agreement with \cite{Andronic:2011yq,Zhao:2011cv,Liu:2009nb}. The reason for this 
discrepancy
is the fact that a different approach for the shadowing factor
for heavy quark production was used in \cite{Capella:2007jv},
which minimized the amount of recombination and led to an overestimation of the total suppression.

Let us finish by an important remark:
we are aware that the comover interaction model at
these energies should not be considered to describe a final-state
interaction at the hadronic level. Indeed, at small values of the proper
time these comovers should be considered as a dense partonic medium. A
large contribution to the comover interaction comes from the few first
fm/c, where the system is in partonic or pre-hadronic stage.
The comover interaction cross section used here, averaged over time,
 do not distinguish between these two scenarios. 
A more refined study would consist on the introduction of different comover interaction cross sections, 
that could vary with the proper time or the densities --using the inverse proportionality between proper time and densities. 
The advantage of the present approach is the economy of parameters and the simplicity of the equations, 
that can be, at least partially, analytically resolved.

\section*{Acknowledgments}
It is a pleasure to thank R. Arnaldi, A. Capella, F. Fleuret, J.-P. Lansberg and  E. Scomparin 
for useful exchanges. 
This work was partially supported by the Ministerio de Ciencia (Spain) \& the IN2P3 (France) (AIC-
D-2011-0740).


\begin{thebibliography}{99}
\bibitem{Matsui86} T.~Matsui, H.~Satz, Phys. Lett. B {\bf 178}, 416
  (1986)

\bibitem{Alessandro05} B.~Alessandro, et al., Eur.\ Phys.\ J.\  C {\bf
    39}, 335 (2005)

\bibitem{Capella97} A.~Capella, A.~Kaidalov, A.~Kouider Akil,
  C.~Gerschel, Phys.\ Lett.\  B {\bf 393}, 431 (1997) 

\bibitem{Armesto98} N.~Armesto, A.~Capella, Phys. Lett. B {\bf
    430}, 23 (1998)

\bibitem{Armesto99} N.~Armesto, A.~Capella, E.G.~Ferreiro, Phys.
  Rev. C {\bf 59}, 395 (1999)

\bibitem{Capella00} A.~Capella, E.G.~Ferreiro, A.B.~Kaidalov,
  Phys. Rev. Lett. {\bf 85}, 2080 (2000)

\bibitem{Thews01} R.~Thews, M.~Schroedter, J.~Rafelski, Phys. Rev.
  C {\bf 63}, 054905 (2001)

\bibitem{Grandchamp02} L.~Grandchamp, R.~Rapp, Nucl. Phys. A {\bf
    709}, 415 (2002)

\bibitem{Yan06} L.~Yan, P.~Zhuang, N.~Xu, Phys.\ Rev.\ Lett.\
  {\bf 97}, 232301 (2006)

\bibitem{Braun00} P.~Braun-Munzinger, J.~Stachel, Phys. Lett. B {\bf
    490}, 196 (2000)

\bibitem{Andronic03} A.~Andronic, P.~Braun-Munzinger, K.~Redlich,
  J.~Stachel, Phys. Lett. B {\bf 571}, 36 (2003)

\bibitem{Kostyuk03} A.P.~Kostyuk, M.I.~Gorenstein, H.~St\"ocker,
  W.~Greiner, Phys. Rev. C {\bf 68}, 041902 (2003)
  
\bibitem{Capella05} A.~Capella, E.G.~Ferreiro, Eur. Phys. J. C {\bf
    42}, 419 (2005)

\bibitem{PHENIX07} A.~Adare, et al., Phys.\ Rev.\ Lett.\  {\bf 98},
  232301 (2007) 

\bibitem{Leitch07} M.J.~Leitch, J.\ Phys.\ G {\bf 34}, S453 (2007)

\bibitem{Capella:2007jv} 
  A.~Capella, L.~Bravina, E.~G.~Ferreiro, A.~B.~Kaidalov, K.~Tywoniuk and E.~Zabrodin,
  Eur.\ Phys.\ J.\ C {\bf 58}, 437 (2008)

\bibitem{Jager74} C.W.~De Jager, H.~De Vries and C.~De Vries, Atom.\
  Data Nucl.\ Data Tabl.\  {\bf 14}, 479 (1974)

%
%

\bibitem{Braun98} M.A.~Braun, C.~Pajares, C.A.~Salgado, N.~Armesto,
  A.~Capella, Nucl. Phys. B {\bf 509}, 357 (1998)

\bibitem{Capella:2006mb}
  A.~Capella and E.~G.~Ferreiro,
  Phys.\ Rev.\ C {\bf 76}, 064906 (2007) 

\bibitem{Lourenco:2008sk} 
  C.~Lourenco, R.~Vogt and H.~K.~Woehri,
  JHEP {\bf 0902}, 014 (2009)

%

\bibitem{Gribov69} V.N.~Gribov, Sov.\ Phys.\ JETP {\bf 29}, 483 (1969);
  {\it ibid.} {\bf 30}, 709 (1970); {\it ibid.} {\bf 26}, 414 (1968)

\bibitem{Schwimmer75} A.~Schwimmer, Nucl.\ Phys.\ B {\bf 94}, 445 (1975)

\bibitem{Capella99} A.~Capella, A.~Kaidalov, J.~Tran Thanh Van,
  Heavy Ion Phys. {\bf 9}, 169 (1999)


\bibitem{Armesto03} N.~Armesto, A.~Capella, A.B.~Kaidalov,
  J.~Lopez-Albacete, C.A.~Salgado, Eur. Phys. J. C {\bf 29}, 531 (2003)


\bibitem{Capella01} A.~Capella, D.~Sousa, Phys. Lett. B {\bf 511},
  185 (2001)

\bibitem{Capella:2011vi} 
  A.~Capella and E.~G.~Ferreiro,
  Eur.\ Phys.\ J.\ C {\bf 72}, 1936 (2012)

\bibitem{Ferreiro:2008wc} 
  E.~G.~Ferreiro, F.~Fleuret, J.~P.~Lansberg and A.~Rakotozafindrabe,
  Phys.\ Lett.\ B {\bf 680}, 50 (2009)

\bibitem{Rakotozafindrabe:2011rw} 
  A.~Rakotozafindrabe, E.~G.~Ferreiro, F.~Fleuret, J.~P.~Lansberg and N.~Matagne,
  Nucl.\ Phys.\ A {\bf 855}, 327 (2011)

\bibitem{Ferreiro:2008td} 
  E.~G.~Ferreiro,
  Contribution to Rencontres de Moriond 2008: QCD and High Energy Interactions,
  arXiv:0805.2753 [hep-ph].

\bibitem{Ferreiro:2009qr} 
  E.~G.~Ferreiro, F.~Fleuret, J.~P.~Lansberg and A.~Rakotozafindrabe,
  Contribution to Rencontres de Moriond 2009: QCD and High Energy Interactions,
  arXiv:0903.4908 [hep-ph].

\bibitem{Vogt:2010aa} 
  R.~Vogt,
  Phys.\ Rev.\ C {\bf 81}, 044903 (2010)


\bibitem{Brodsky88} S.~Brodsky, A.H.~Mueller, Phys. Lett. B {\bf
    206}, 685 (1988)
 
\bibitem{Koch90} B.~Koch, U.~Heinz, J.~Pitsut, Phys. Lett. {\bf
    243}, 149 (1990)

\bibitem{Capella95} A.~Capella, Phys.\ Lett.\  B {\bf 364}, 175 (1995)

\bibitem{Capella96} A.~Capella, A.~Kaidalov, A.~Kouider Akil,
  C.~Merino, J.~Tran Thanh Van, Z.\ Phys.\  C {\bf 70}, 507 (1996)\\
  A.~Capella, C.A.~Salgado, D.~Sousa, Eur.\ Phys.\ J.\ C {\bf 30},
  111 (2003)


\bibitem{Abelev:2012kr} 
  B.~Abelev {\it et al.}  [ALICE Collaboration],
Phys.\ Lett.\ B {\bf 718}, 295 (2012)

\bibitem{Bossu:2011qe} 
  F.~Bossu, Z.~C.~del Valle, A.~de Falco, M.~Gagliardi, S.~Grigoryan and G.~Martinez Garcia,
  arXiv:1103.2394 [nucl-ex].

\bibitem{:2012sx} 
  B.~Abelev [ALICE Collaboration],
  JHEP {\bf 1207}, 191 (2012)


\bibitem{Abelev:2012rv} 
  B.~Abelev {\it et al.}  [ALICE Collaboration],
  Phys.\ Rev.\ Lett.\  {\bf 109}, 072301 (2012)

\bibitem{Wiechula:2012mh} 
  J.~Wiechula [for the ALICE Collaboration],
  arXiv:1208.6566 [hep-ex].

\bibitem{ScomparinQM2012} 
E.~Scomparin [ALICE Collaboration],
  Nucl.\ Phys.\ A904-905 {\bf 2013}, 202c (2013);
I.~-C.~Arsene [ALICE Collaboration],
  Nucl.\ Phys.\ A904-905 {\bf 2013}, 623c (2013)

\bibitem{ArnaldiQM2012}
R.~Arnaldi [ALICE Collaboration],
  Nucl.\ Phys.\ A {\bf 904-905}, 595c (2013)


\bibitem{Zhao:2011cv} 
  X.~Zhao and R.~Rapp,
  Nucl.\ Phys.\ A {\bf 859}, 114 (2011)

\bibitem{Andronic:2011yq} 
  A.~Andronic, P.~Braun-Munzinger, K.~Redlich and J.~Stachel,
  J.\ Phys.\ G G {\bf 38}, 124081 (2011)

\bibitem{Ferreiro:2013pua} 
  E.~G.~Ferreiro, F.~Fleuret, J.~P.~Lansberg and A.~Rakotozafindrabe,
  Phys.\ Rev.\ C {\bf 88}, 047901 (2013)

\bibitem{Capella:2006fw}
  A.~Capella and E.~G.~Ferreiro,
  Phys.\ Rev.\ C {\bf 75}, 024905 (2007) 

\bibitem{CapFerr2014}
A.~Capella and E.~G.~Ferreiro, in preparation.

\bibitem{Chatrchyan:2012np} 
  S.~Chatrchyan {\it et al.}  [CMS Collaboration],
  JHEP {\bf 1205}, 063 (2012)

\bibitem{MironovQM2012}
C. Mironov [for the CMS Collaboration],
Contribution to Quark Matter 2012,
  Nucl.\ Phys.\ A904-905 {\bf 2013}, 194c (2013)

\bibitem{MoonQM2012}
D. H. Moon  [for the CMS Collaboration],
Contribution to Quark Matter 2012,
  Nucl.\ Phys.\ A904-905 {\bf 2013}, 591c (2013)

\bibitem{Liu:2009nb} 
  Y.~-p.~Liu, Z.~Qu, N.~Xu and P.~-f.~Zhuang,
  Phys.\ Lett.\ B {\bf 678}, 72 (2009)

\end{thebibliography}
\end{document}